\documentclass[twocolumn,letterpaper,pra]{revtex4}
\usepackage{graphicx}% eps files
\usepackage{bm}% bold math
\usepackage{amsmath}% equation formatting
\usepackage{amssymb}% math symbols
\newlength{\twocolumnwidth}

\begin{document} 
%******************************************* 
\title{S-ordered phase-space path integrals and time-s-ordering of Heisenberg operators\\ (reseach notes)} 
%*******************************************
\author{L.\ I.\ Plimak} 
\affiliation{Abteilung Quantenphysik, Universit\"at Ulm, 
D-89069 Ulm, Germany} 
%******************************************* 
%\date{\today} 
%*******************************************
\begin{abstract} 
%*******************************************
Formal structure of phase-space path integrals based on different types of operator orderings is analysed. 
%******************************************* 
\end{abstract}
%******************************************* 
%\pacs{XXZ}
%*******************************************
\maketitle 
%******************************************* 
\section{Introductory remarks} 
%******************************************* 
Section \ref{ch:W} of these notes was written as part of my discussion of phase-space path-integral methods with \mbox{A.\ Polkovnikov}. I wanted to translate his approach \cite{Polkan} for myself into a familiar language. Among other things it is shown that if we construct a phase-space path integral using the symmetric ordering of free-field operators, the path integral expresses naturally quantum averages of time-symmetrically ordered products of Heisenberg operators. These resuls were included, with minor changes, into our paper \cite{Prep}. Later I realised that the resuls may be generalised to an arbitrary ordering of free-field operators. As is shown in section \ref{ch:S}, for any ordering of free-field operators one finds the corresponding ordering of Heisenberg operators. In particular, using the normal ordering for the path integral results in Glauber-Kelly-Kleiner's time-normal ordering of Heisenberg operators. The fact that commutators of Heisenberg operators for different times are related to response properties of the system also holds irrespective of the choice of the underlying operators ordering. 

At the time of writing these notes, I found it easier to make as much use as possible of the Weyl-based approach also in the case of a non-Weyl ordering. A much better idea would be treating all orderings on an equal basis, but this belongs in the future. In the meantime I decided to post the notes as they are. Perhaps I am not the only one who finds the results interesting. 
%******************************************* 
\section{Multitime Wigner representation}\label{ch:W}
%******************************************* 
\subsection{Preliminaries} 
%******************************************* 
We firstly refresh our memory on the Wigner representation. We follow the paper by Agarwal and Wolf \cite{AgarwalWolf} (perhaps more in spirit than to the letter). A complex function $A(\alpha)$ is the symmetric, or Wigner, representation of the operator $\hat A$ if 
%=============================================
{\begin{multline}\hspace{0.4\columnwidth}\hspace{-0.4\twocolumnwidth} 
\hat A = \int \frac{d^2 \alpha}{\pi }A(\alpha)\int \frac{d^2 \beta }{\pi }\text{e}^{
\beta (\hat a - \alpha)^{\dag}
-\beta^* (\hat a - \alpha)}
\\ 
= \int \frac{d^2 \alpha\, d^2 \beta }{\pi ^2}\text{e}^{
\alpha\beta^* - \alpha^*\beta }A(\alpha) \hat D(\beta ), 
\hspace{0.4\columnwidth}\hspace{-0.4\twocolumnwidth} 
% \nonumber 
\label{eq:DefW} 
\end{multline}}%
%+++++++++++++++++++++++++++++++++++++++++++++
where $\hat D(\beta )$ is the complex displacement operator, 
%=============================================
\protect{\begin{align}{{
 \begin{aligned} 
\hat D(\beta ) & = \text{e}^{
\beta \hat a^{\dag}
-\beta^* \hat a }, & \text{Tr} \hat D^{\dag}(\beta)\hat D(\beta') = \pi \delta^{(2)}(\beta -\beta '). 
\end{aligned}}}
% \nonumber % \eqlabel{} 
\end{align}}%
%+++++++++++++++++++++++++++++++++++++++++++++
Equation (\ref{eq:DefW}) may be obtained by using completeness of the set of displacement operators, 
%=============================================
\protect{\begin{align}{{
 \begin{aligned} 
\hat A = \int \frac{d^2 \beta }{\pi } \hat D(\beta )\text{Tr} \hat A \hat D^{\dag}(\beta ), 
\end{aligned}}}
% \nonumber % \eqlabel{} 
\end{align}}%
%+++++++++++++++++++++++++++++++++++++++++++++
so that 
%=============================================
\protect{\begin{align}{{
 \begin{aligned} 
A(\alpha) = \protect\protect\ensuremath{\big[
\hat A
\big]} (\alpha) = \int \frac{d^2 \beta }{\pi }\text{e}^{
\beta\alpha^* - \beta^*\alpha }\text{Tr} \hat A \hat D^{\dag}(\beta ) . 
\end{aligned}}}
% \nonumber 
\label{eq:EqW} 
\end{align}}%
%+++++++++++++++++++++++++++++++++++++++++++++
The alternative notation $\protect\protect\ensuremath{\big[
\hat A
\big]} (\alpha)$ for the Wigner representation of $\hat A$ comes especially handy when $\hat A$ is an operator expression. 
Using that $\hat D^{\dag}(\beta ) = \hat D(-\beta )$ it is easy to prove the correspondence rule for the trace, 
%=============================================
\protect{\begin{align}{{
 \begin{aligned} 
\text{Tr} \hat A \hat B = \int \frac{d^2 \alpha}{\pi } A(\alpha)B(\alpha) . 
\end{aligned}}}
% \nonumber 
\label{eq:TrAB} 
\end{align}}%
%+++++++++++++++++++++++++++++++++++++++++++++

The symmetric representation of the rho-matrix $\hat \rho $ is called the Wigner function and often denoted $W(\alpha)$. However we follow our general pattern of using the same letter for the operator and its symmetric representation and denote the Wigner function as $\rho (\alpha)$. 

All calculations in this paper are in essence based on one key formula expressing the Wigner representation of an operator product $\hat A \hat B$, denoted as $[\hat A\hat B](\alpha)$, by the Wigner representations of the factors. Such relation follows by firstly expressing the operators by their symmetric representations using eq.\ (\protect\ref{eq:DefW}), and then employing eq.\ (\protect\ref{eq:EqW}) to express $[\hat A\hat B](\alpha)$. It is easy to verify that 
%=============================================
\protect{\begin{align}{{
 \begin{aligned} 
\hat D(\beta ) \hat D(\beta ') = \hat D(\beta +\beta ')
\text{e}^{\frac{1}{2}(\beta \beta ^{\prime *} - \beta^* \beta' ) } . 
\end{aligned}}}
% \nonumber % \eqlabel{} 
\end{align}}%
%+++++++++++++++++++++++++++++++++++++++++++++
The rest of the calculation is trivial algebra. As a result we obtain 
%=============================================
{\begin{multline}\hspace{0.4\columnwidth}\hspace{-0.4\twocolumnwidth} 
{[\hat A\hat B]}(\alpha) = \int \frac{d^2 \alpha _0 d^2\sigma }{\pi ^2} 
\text{e}^{(\alpha -\alpha _0)\sigma ^* - (\alpha -\alpha _0)^*\sigma } 
\\ \times 
A(\alpha_0 ) 
B(\alpha_0 + \sigma /2) . 
\hspace{0.4\columnwidth}\hspace{-0.4\twocolumnwidth} 
% \nonumber 
\label{eq:AB1} 
\end{multline}}%
%+++++++++++++++++++++++++++++++++++++++++++++
By the change of variable $\alpha _0\to\alpha _0 + \sigma /2$ we can write it in the alternative form 
%=============================================
{\begin{multline}\hspace{0.4\columnwidth}\hspace{-0.4\twocolumnwidth} 
{[\hat A\hat B]}(\alpha) = \int \frac{d^2 \alpha _0 d^2\sigma }{\pi ^2} 
\text{e}^{(\alpha -\alpha _0)\sigma ^* - (\alpha -\alpha _0)^*\sigma } 
\\ \times 
A(\alpha_0 - \sigma /2) 
B(\alpha_0 ) . 
\hspace{0.4\columnwidth}\hspace{-0.4\twocolumnwidth} 
% \nonumber 
\label{eq:AB2} 
\end{multline}}%
%+++++++++++++++++++++++++++++++++++++++++++++
The usual operator-to-phase-space correspondences may be obtained from eqs.\ (\protect\ref{eq:AB1}) and (\ref{eq:AB2}). For instance, from eq.\ (\protect\ref{eq:AB1})
\begin{widetext} 
%=============================================
{\begin{multline} 
{[\hat \rho \hat a]}(\alpha) = \int 
\frac{d^2 \alpha _0 d^2\sigma }{\pi ^2} 
\text{e}^{(\alpha -\alpha _0)\sigma ^* - (\alpha -\alpha _0)^*\sigma }\rho(\alpha _0 )(\alpha _0 + \sigma /2) 
\\ = \int \frac{d^2 \alpha _0 d^2\sigma }{\pi ^2} 
 \rho(\alpha_0 )
\protect\ensuremath{\Big(
\alpha_0 + \frac{1}{2} \frac{\partial }{\partial \alpha _0 ^*} \Big)}
\text{e}^{(\alpha -\alpha _0)\sigma ^* - (\alpha -\alpha _0)^*\sigma } . 
% \nonumber % \eqlabel{} 
\end{multline}}%
%+++++++++++++++++++++++++++++++++++++++++++++
\end{widetext}%
Integrating by parts we move ${\partial }/{\partial \alpha _0 ^*}$ on \mbox{$\rho(\alpha_0)$}. In particular this ``frees'' the integration over $\sigma $, 
%=============================================
\protect{\begin{align}{{
 \begin{aligned} 
\int \frac{d^2 \sigma }{\pi } 
\text{e}^{(\alpha -\alpha _0)\sigma ^* - (\alpha -\alpha _0)^*\sigma } = 
\pi \delta^{(2)}(\alpha -\alpha _0), 
\end{aligned}}}
% \nonumber 
\label{eq:DAA} 
\end{align}}%
%+++++++++++++++++++++++++++++++++++++++++++++
so that the remaining integral over $\alpha _0$ is taken trivially. We have thus recovered the first of the four standard correpondences characteristic of the Wigner representation, 
%=============================================
\protect{\begin{gather}{{
 \begin{gathered} 
\hat \rho \hat a \Longleftrightarrow \protect\ensuremath{\Big(
\alpha - \frac{1}{2} \frac{\partial }{\partial \alpha ^*} \Big)} 
\rho(\alpha), 
\\ 
\hat a \hat \rho \Longleftrightarrow \protect\ensuremath{\Big(
\alpha + \frac{1}{2} \frac{\partial }{\partial \alpha ^*} \Big)} 
\rho(\alpha), 
\\ 
\hat \rho \hat a^{\dag} \Longleftrightarrow \protect\ensuremath{\Big(
\alpha ^* + \frac{1}{2} \frac{\partial }{\partial \alpha} \Big)} 
\rho(\alpha), 
\\ 
\hat a^{\dag} \hat \rho \Longleftrightarrow \protect\ensuremath{\Big(
\alpha ^* - \frac{1}{2} \frac{\partial }{\partial \alpha} \Big)} 
\rho(\alpha). 
\end{gathered}}}
% \nonumber 
\label{eq:Corr} 
\end{gather}}%
%+++++++++++++++++++++++++++++++++++++++++++++
The other three follow with equal ease. With $\rho (\alpha)\to A(\alpha)$ they hold not only for the rho-matrix but for any operator $\hat A$. 
%******************************************* 
\subsection{Phase-space transition amplitude} 
%******************************************* 
We assume that the coincidence point for the Schr\"odinger and Heisenberg pictures is at $t=t_0$. The Heisenberg rho-matrix $\hat \rho $ then equals with the Schr\"odinger rho-matrix $\hat \rho (t)$ at $t=t_0$, 
%=============================================
\protect{\begin{align}{{
 \begin{aligned} 
\hat \rho = \hat \rho (t_0) . 
\end{aligned}}}
% \nonumber % \eqlabel{} 
\end{align}}%
%+++++++++++++++++++++++++++++++++++++++++++++
It is convenient to preserve $t_0$ in the notation: the Wigner function of the Heisenberg rho-matrix will be denoted $\rho (\alpha,t_0)$. Consider the quantum average of a Heisenberg operator $\protect{\hat{\mathcal B}}(t)$, $t>t_0$ 
%=============================================
\protect{\begin{align}{{
 \begin{aligned} 
\protect\protect\ensuremath{\big\langle 
\protect{\hat{\mathcal B}}(t)
\big\rangle} = \text{Tr} \protect{\hat{\mathcal B}}(t)\hat \rho (t_0) . 
\end{aligned}}}
% \nonumber 
\label{eq:BAv0} 
\end{align}}%
%+++++++++++++++++++++++++++++++++++++++++++++
The expression for $\protect{\hat{\mathcal B}}(t)$ in terms of the evolution operator, 
%=============================================
\protect{\begin{align}{{
 \begin{aligned} 
\protect{\hat{\mathcal U}}(t,t_0) = \text{e}^{-i(t-t_0)\hat H/\hbar }, 
\end{aligned}}}
% \nonumber 
\label{eq:UStat} 
\end{align}}%
%+++++++++++++++++++++++++++++++++++++++++++++
is 
%=============================================
\protect{\begin{align}{{
 \begin{aligned} 
\protect{\hat{\mathcal B}}(t) = \protect{\hat{\mathcal U}}^{\dag}(t,t_0)\hat B(t)\protect{\hat{\mathcal U}}(t,t_0) , 
\end{aligned}}}
% \nonumber 
\label{eq:Bt} 
\end{align}}%
%+++++++++++++++++++++++++++++++++++++++++++++
where $\hat B(t)$ is $\protect{\hat{\mathcal B}}(t)$ in the Schr\"odinger picture (time-dependent, in general). By making use of the equations (\ref{eq:DefW})--(\ref{eq:TrAB}) we write 
%=============================================
\protect{\begin{align}{{
 \begin{aligned} 
\protect\protect\ensuremath{\big\langle 
\protect{\hat{\mathcal B}}(t)
\big\rangle} = \int \frac{d^2 \alpha _0d^2 \alpha }{\pi^2} 
B(\alpha ,t)U(\alpha ,t,\alpha _0,t_0)\rho (\alpha,t_0), 
\end{aligned}}}
% \nonumber 
\label{eq:BAv} 
\end{align}}%
%+++++++++++++++++++++++++++++++++++++++++++++
where $B(\alpha ,t) = \protect\protect\ensuremath{\big[
\hat B(t)
\big]} (\alpha)$. The {\em phase-space transition amplitude\/} $U(\alpha ,t,\alpha _0,t_0)$ is defined by 
%=============================================
{\begin{multline}\hspace{0.4\columnwidth}\hspace{-0.4\twocolumnwidth} 
U(\alpha ,t,\alpha _0,t_0) = \int \frac{d^2 \beta _0d^2 \beta }{\pi^2}\text{e}^{
\alpha\beta^* - \alpha^*\beta + \alpha_0\beta_0^* - \alpha_0^*\beta_0}
\\ \times 
\text{Tr}\hat D^{\dag}(\beta_0 )\protect{\hat{\mathcal U}}^{\dag}(t,t_0)\hat D^{\dag}(\beta )\protect{\hat{\mathcal U}}(t,t_0)
\hspace{0.4\columnwidth}\hspace{-0.4\twocolumnwidth} 
% \nonumber 
\label{eq:UDef} 
\end{multline}}%
%+++++++++++++++++++++++++++++++++++++++++++++
For $U(\alpha ,t,\alpha _0,t_0)$ we have natural conditions 
%=============================================
\protect{\begin{align}{{
 \begin{aligned} 
U(\alpha ,t_0,\alpha _0,t_0) = U(\alpha ,t,\alpha _0,t_0)|_{\protect{\hat{\mathcal H}} = 0} =
\pi \delta ^{(2)}(\alpha -\alpha _0). 
\end{aligned}}}
% \nonumber 
\label{eq:Dlt} 
\end{align}}%
%+++++++++++++++++++++++++++++++++++++++++++++
Unitarity of the quantum evolution
is expressed in terms of $U(\alpha ,t,\alpha _0,t_0)$ as 
%=============================================
\protect{\begin{align}{{
 \begin{aligned} 
\int \frac{d^2 \alpha}{\pi }U(\alpha ,t,\alpha _0,t_0) = 1 . 
\end{aligned}}}
% \nonumber % \eqlabel{} 
\end{align}}%
%+++++++++++++++++++++++++++++++++++++++++++++
This property is easily proven using eq.\ (\protect\ref{eq:UDef}) and unitarity of $\protect{\hat{\mathcal U}}(t,t_0)$. Furthermore, the group property of the evolution operator, ($t>t_1>t_2$) 
%=============================================
\protect{\begin{align}{{
 \begin{aligned} 
\protect{\hat{\mathcal U}}(t,t_0) = \protect{\hat{\mathcal U}}(t,t_1)\protect{\hat{\mathcal U}}(t_1,t_0), 
\end{aligned}}}
% \nonumber % \eqlabel{} 
\end{align}}%
%+++++++++++++++++++++++++++++++++++++++++++++
results in the group property of the phase-space amplitude, 
%=============================================
\protect{\begin{align}{{
 \begin{aligned} 
U(\alpha ,t,\alpha _0,t_0) = \int \frac{d^2 \alpha_1}{\pi }U(\alpha ,t,\alpha _1,t_1)U(\alpha_1 ,t_1,\alpha _0,t_0) . 
\end{aligned}}}
% \nonumber 
\label{eq:UU} 
\end{align}}%
%+++++++++++++++++++++++++++++++++++++++++++++

Some calculations in the below are much simplified if we associate the group property (\ref{eq:UU}) with the change of the coincidence point $t_0\to t_1$. Namely, the Schr\"odinger picture stays put while there are many Heisenberg pictures parametrised by $t_0$ in eqs.\ (\protect\ref{eq:UStat}) and (\ref{eq:Bt}). To show the dependence of a Heisenberg operator on the coincidence point $t_1$ we shall denote it as $\protect{\hat{\mathcal B}}_{t_1}(t)$ (say). The absence of index signals the standard coincidence point, $\protect{\hat{\mathcal B}}(t)=\protect{\hat{\mathcal B}}_{t_0}(t)$. 

Assume now the group property is inserted in eq.\ (\protect\ref{eq:BAv}). We immediately recognise 
%=============================================
{\begin{multline}\hspace{0.4\columnwidth}\hspace{-0.4\twocolumnwidth} 
\int \frac{d^2 \alpha_0}{\pi }U(\alpha_1 ,t_1,\alpha _0,t_0)\rho (\alpha _0,t_0) \\ = 
\rho (\alpha_1 ,t_1) = 
\protect\protect\ensuremath{\big[
\hat \rho (t_1)
\big]} (\alpha_1)
\hspace{0.4\columnwidth}\hspace{-0.4\twocolumnwidth} 
% \nonumber 
\label{eq:Rt1} 
\end{multline}}%
%+++++++++++++++++++++++++++++++++++++++++++++
as the Wigner representation of the Schr\"odinger rho-matrix $\rho (\alpha_1 ,t_1)$, which is at the same time the Wigner representation of the Heisenberg rho-matrix with coincidence point $t_1$. We have then to conclude that 
%=============================================
\protect{\begin{align}{{
 \begin{aligned} 
\int \frac{d^2 \alpha}{\pi }B(\alpha,t)U(\alpha ,t,\alpha _1,t_1) = 
\protect\protect\ensuremath{\big[
\protect{\hat{\mathcal B}}_{t_1}(t)
\big]} (\alpha_1), 
\end{aligned}}}
% \nonumber 
\label{eq:BU} 
\end{align}}%
%+++++++++++++++++++++++++++++++++++++++++++++
as a consequence of (\ref{eq:Rt1}) and of the relation 
%=============================================
\protect{\begin{align}{{
 \begin{aligned} 
\protect\protect\ensuremath{\big\langle 
\protect{\hat{\mathcal B}}(t)
\big\rangle} = \text{Tr}\protect{\hat{\mathcal B}}_{t_1}(t)\hat \rho (t_1)
\end{aligned}}}
% \nonumber % \eqlabel{} 
\end{align}}%
%+++++++++++++++++++++++++++++++++++++++++++++
expressing preservation of the averages. Using (\ref{eq:Bt}) to express $\protect{\hat{\mathcal B}}_{t_1}(t)$ by $\protect{\hat{\mathcal B}}(t)$ we find a useful relation 
%=============================================
{\begin{multline}\hspace{0.4\columnwidth}\hspace{-0.4\twocolumnwidth} 
\int \frac{d^2 \alpha}{\pi }B(\alpha,t)U(\alpha ,t,\alpha _1,t_1) \\ = 
\protect\protect\ensuremath{\big[
\protect{\hat{\mathcal U}}(t_1,t_0)\protect{\hat{\mathcal B}}(t)\protect{\hat{\mathcal U}}^{\dag}(t_1,t_0)
\big]} (\alpha_1). 
\hspace{0.4\columnwidth}\hspace{-0.4\twocolumnwidth} 
% \nonumber 
\label{eq:Usf} 
\end{multline}}%
%+++++++++++++++++++++++++++++++++++++++++++++

%******************************************* 
\subsection{Phase-space path integral and the truncated Wigner representation} 
%******************************************* 
By itself, the amplitude $U(\alpha ,t,\alpha _0,t_0)$ is not associated with a path-integral approach. In this respect it is similar to the coordinate-space transition amplitude, of which Feynman's path-integral representation is only one of the multitude of ways of approaching it. However, unlike the coordinate-space transition amplitude, a path-integral representation of the phase-space amplitude is a natural --- and practical --- way of looking at the phase-space evolution. 

While for simplicity we assumed in eq.\ (\protect\ref{eq:UStat}) that the Hamiltonian is time-independent, the definition (\ref{eq:UDef}) does not make use of this assumption, nor does our notation. If the Schr\"odinger Hamiltonian $\hat H(t)$ is time dependent, in place of (\ref{eq:UStat}) one has to use the Schr\"odinger equation for 
$\protect{\hat{\mathcal U}}(t,t_0)$, 
%=============================================
\protect{\begin{align}{{
 \begin{aligned} 
i\hbar {\dot{\protect{\hat{\mathcal U}}}}{(t,t_0)} & = \hat H(t)\protect{\hat{\mathcal U}}(t,t_0), & 
\protect{\hat{\mathcal U}}(t_0,t_0) = \hat{\openone} . 
\end{aligned}}}
% \nonumber % \eqlabel{} 
\end{align}}%
%+++++++++++++++++++++++++++++++++++++++++++++
For infinitesimally small time interval $t-t_0=\Delta t$ the evolution operator is given by the approximate formula 
%=============================================
\protect{\begin{align}{{
 \begin{aligned} 
\protect{\hat{\mathcal U}}(t_0+\Delta t,t_0)\approx 
\hat{\openone} - i \hat H(t_0)\Delta t/\hbar . 
\end{aligned}}}
% \nonumber % \eqlabel{} 
\end{align}}%
%+++++++++++++++++++++++++++++++++++++++++++++
One can then construct the evolution operator for finite time intervals as an infinite product of the infinitesimal evolution operators. By virtue of the group property (\ref{eq:UU}) this automatically expresses the phase-space transition amplitude as a path integral: 
\begin{widetext} 
%=============================================
\protect{\begin{align}{{
 \begin{aligned} 
U(\alpha ,t,\alpha _0,t_0) = \lim_{N\to \infty}\int 
U(\alpha ,t,\alpha _N,t_N)
%\\ \times 
\prod_{k=1}^N\frac{d^2\alpha_k 
U(\alpha_k ,t_k,\alpha _{k-1},t_{k-1})}{\pi } , 
\end{aligned}}}
% \nonumber 
\label{eq:PI} 
\end{align}}%
%+++++++++++++++++++++++++++++++++++++++++++++
where we have assumed that the interval $t-t_0$ is sliced into $N+1$ infinitesimal intervals $\Delta t = (t-t_0)/(N+1)$, and $t_k = t_0 + k\Delta t, \ k = 0,\cdots,N$ are their left ends. 

Unlike the operators, all factors in (\ref{eq:PI}) are commuting c-numbers so that the properties of the path integral can be fully understood from the properties of the transition amplitude over an infinitesimal time interval $U(\alpha ,t+\Delta t,\alpha _0,t)$. 
To find it, consider the evolution of the Schr\"odinger rho-matrix from $t$ to $t+\Delta t$: 
%=============================================
\protect{\begin{align}{{
 \begin{aligned} 
\hat \rho (t+\Delta t) = \hat \rho (t) - i[\hat H(t)\hat \rho (t)-\hat \rho (t)\hat H(t)]\Delta t/\hbar . 
\end{aligned}}}
% \nonumber 
\label{eq:drho} 
\end{align}}%
%+++++++++++++++++++++++++++++++++++++++++++++
By making use of eqs.\ (\protect\ref{eq:AB1}) and (\ref{eq:AB2}) we can turn this into a relation between the corresponding Wigner functions: 
%=============================================
{\begin{multline} 
\rho (\alpha,t+\Delta t) = 
\int \frac{d^2\alpha _0 d^2 \sigma }{\pi ^2} 
\text{e}^{(\alpha -\alpha _0)\sigma ^* - (\alpha -\alpha _0)^*\sigma } 
\rho (\alpha_0,t) 
%\\ \times 
\protect\protect\ensuremath{\bigg\{
1 - \frac{i\Delta t}{\hbar } \protect\protect\ensuremath{\Big[
H(\alpha_0 - \sigma /2,t) - H(\alpha_0 + \sigma /2,t)
\Big]} 
\bigg\}} , 
% \nonumber 
\label{eq:Urho0} 
\end{multline}}%
%+++++++++++++++++++++++++++++++++++++++++++++
where $H(\alpha ,t)$ is the Wigner representation of the Hamitonian. The zeroth-order term was dragged under the integral using eq.\ (\protect\ref{eq:DAA}). To recover \mbox{$U(\alpha ,t+\Delta t,\alpha _0,t)$} from this formula we note that it follows from (\ref{eq:BAv}) that 
%=============================================
\protect{\begin{align}{{
 \begin{aligned} 
\rho (\alpha,t+\Delta t) = \int \frac{d^2 \alpha_0}{\pi }
U(\alpha ,t+\Delta t,\alpha _0,t)\rho (\alpha_0,t). 
\end{aligned}}}
% \nonumber 
\label{eq:drho1} 
\end{align}}%
%+++++++++++++++++++++++++++++++++++++++++++++
This way, 
%=============================================
\protect{\begin{align}{{
 \begin{aligned} 
U(\alpha ,t+\Delta t,\alpha _0,t) = \int \frac{d^2 \sigma }{\pi }\exp\protect\protect\ensuremath{\bigg\{
(\alpha -\alpha _0)\sigma ^* - (\alpha -\alpha _0)^*\sigma - \frac{i\Delta t}{\hbar } \protect\protect\ensuremath{\Big[
H(\alpha_0 - \sigma /2,t) - H(\alpha_0 + \sigma /2,t)
\Big]}
\bigg\}} . 
\end{aligned}}}
% \nonumber 
\label{eq:UInf} 
\end{align}}%
%+++++++++++++++++++++++++++++++++++++++++++++
In this formula, we have also replaced the expression in the curly brackets in (\ref{eq:Urho0}) by the exponent. This is justified for an infinitesimally small $\Delta t$. 

Following \cite{Polkan} we separate in (\ref{eq:UInf}) the classical evolution from quantum fluctuations. Formally, we write 
%=============================================
\protect{\begin{align}{{
 \begin{aligned} 
H(\alpha_0 + \sigma /2,t) - H(\alpha_0 - \sigma /2,t) = \sigma f^*(\alpha_0,t) + \sigma ^* f(\alpha_0,t) + h^{(3)}(\alpha_0,\sigma,t), 
\end{aligned}}}
% \nonumber % \eqlabel{} 
\end{align}}%
%+++++++++++++++++++++++++++++++++++++++++++++
where 
%=============================================
\protect{\begin{align}{{
 \begin{aligned} 
f(\alpha_0,t) &= \frac{\partial H(\alpha_0 ,t)}{\partial \alpha_0^*}, & 
f^*(\alpha_0,t) &= \frac{\partial H(\alpha_0 ,t)}{\partial \alpha_0}. 
\end{aligned}}}
% \nonumber % \eqlabel{} 
\end{align}}%
%+++++++++++++++++++++++++++++++++++++++++++++
This way, 
%=============================================
\protect{\begin{align}{{
 \begin{aligned} 
U(\alpha ,t+\Delta t,\alpha _0,t) = \int \frac{d^2 \sigma }{\pi }\exp\protect\protect\ensuremath{\bigg\{
(\alpha -\alpha _0 + if\Delta t /\hbar )\sigma ^* - (\alpha -\alpha _0 + if\Delta t /\hbar )^*\sigma + \frac{i\Delta t}{\hbar }h^{(3)}(\alpha_0,\sigma,t) 
\bigg\}} . 
\end{aligned}}}
% \nonumber 
\label{eq:UInf1} 
\end{align}}%
%+++++++++++++++++++++++++++++++++++++++++++++
For polynomial Hamitonians $h^{(3)}(\alpha_0,\sigma,t)$ is a polynomial where all terms are at least cubic in $\sigma ,\sigma ^*$. Nonzero $h^{(3)}(\alpha_0,\sigma,t)$ can only occur for nonlinear interactions. If for some physical reason $h^{(3)}$ can be neglected, then 
%=============================================
\protect{\begin{align}{{
 \begin{aligned} 
U(\alpha ,t+\Delta t,\alpha _0,t) = \pi \delta ^{(2)}(\alpha -\alpha _0 + if \Delta t/h) . 
\end{aligned}}}
% \nonumber % \eqlabel{} 
\end{align}}%
%+++++++++++++++++++++++++++++++++++++++++++++
In this case the motion in phase space is along the trajectories obeying the equation 
%=============================================
\protect{\begin{align}{{
 \begin{aligned} 
i\hbar \dot \alpha = f(\alpha,t). 
\end{aligned}}}
% \nonumber % \eqlabel{} 
\end{align}}%
%+++++++++++++++++++++++++++++++++++++++++++++
This motion is deterministic (nonstochastic) and in this sense classical \cite{But2}. Nonzero $h^{(3)}$ introduces quantum fluctuations \cite{But} which may be accounted for exactly \cite{3on} or perturbatively \cite{Polkan}. %In this paper we assume that $h^{(3)}$ may be neglected. 
%******************************************* 
\subsection{Two-time averages} 
%******************************************* 
Consider the quantum average ($t_0<t_1,t_2$)
%=============================================
\protect{\begin{align}{{
 \begin{aligned} 
\protect\protect\ensuremath{\big\langle 
\protect{\hat{\mathcal X}}(t_1)\protect{\hat{\mathcal Y}}(t_2)
\big\rangle} = \text{Tr}\hat \rho (t_0)\protect{\hat{\mathcal X}}(t_1)\protect{\hat{\mathcal Y}}(t_2) , 
\end{aligned}}}
% \nonumber % \eqlabel{} 
\end{align}}%
%+++++++++++++++++++++++++++++++++++++++++++++
where $\protect{\hat{\mathcal X}}(t)$ and $\protect{\hat{\mathcal Y}}(t)$ are Heisenberg operators. The evolution operator relates them to their Schr\"odinger representations $\hat X, \hat Y$, cf.\ eq.\ (\protect\ref{eq:Bt}). 
Rather than distinguishing the cases $t_1>t_2$ and $t_1<t_2$, we assume that $t_1<t_2$ and consider two distinct averages, $\protect\protect\ensuremath{\big\langle 
\protect{\hat{\mathcal X}}(t_1)\protect{\hat{\mathcal Y}}(t_2)
\big\rangle}$ and $\protect\protect\ensuremath{\big\langle 
\protect{\hat{\mathcal Y}}(t_2)\protect{\hat{\mathcal X}}(t_1)
\big\rangle}$. Simple formulae may be found for the practically important case $\protect{\hat{\mathcal X}}(t)=\protect{\hat{\mathcal A}}(t),\protect{\hat{\mathcal A}}^{\dag}(t)$, where 
%=============================================
\protect{\begin{align}{{
 \begin{aligned} 
\protect{\hat{\mathcal A}}(t) & = \protect{\hat{\mathcal U}}^{\dag} (t,t_0)\hat a\protect{\hat{\mathcal U}} (t,t_0), & 
\protect{\hat{\mathcal A}}^{\dag}(t) & = \protect{\hat{\mathcal U}}^{\dag} (t,t_0)\hat a^{\dag}\protect{\hat{\mathcal U}} (t,t_0), 
\end{aligned}}}
% \nonumber 
\label{eq:HeisA} 
\end{align}}%
%+++++++++++++++++++++++++++++++++++++++++++++
are the Heisenberg field operators. Consider, for example, the average $\protect\protect\ensuremath{\big\langle 
\protect{\hat{\mathcal Y}}(t_2)\protect{\hat{\mathcal A}}(t_1)
\big\rangle}$. Moving the coincidence point to $t=t_1$ and using eq.\ (\protect\ref{eq:Usf}) we have 
%=============================================
\protect{\begin{align}{{
 \begin{aligned} 
\protect\protect\ensuremath{\big\langle 
\protect{\hat{\mathcal Y}}(t_2)\protect{\hat{\mathcal A}}(t_1)
\big\rangle} = \text{Tr} \protect{\hat{\mathcal Y}}_{t_1}(t_2)\hat a \hat\rho (t_1) = %\\ 
\int \frac{d^2 \alpha_1d^2 \alpha_2}{\pi^2 }Y(\alpha_2)U(\alpha_2,t_2,\alpha_1,t_1)
\protect\protect\ensuremath{\big[
\hat a \hat \rho (t_1) 
\big]} (\alpha_1). 
\end{aligned}}}
% \nonumber % \eqlabel{} 
\end{align}}%
%+++++++++++++++++++++++++++++++++++++++++++++
The phase-space correspondences (\ref{eq:Corr}) allow us to write 
%=============================================
\protect{\begin{align}{{
 \begin{aligned} 
\protect\protect\ensuremath{\big[
\hat a \hat \rho (t_1) 
\big]} (\alpha_1) = \protect\ensuremath{\Big(
\alpha_1 + \frac{1}{2} \frac{\partial }{\partial \alpha_1 ^*} \Big)} 
\rho (\alpha_1,t_1). 
\end{aligned}}}
% \nonumber % \eqlabel{} 
\end{align}}%
%+++++++++++++++++++++++++++++++++++++++++++++
where $\rho (\alpha_1,t_1)$ is given by eq.\ (\protect\ref{eq:Rt1}). As a result of these manipulations we obtain 
%=============================================
\protect{\begin{align}{{
 \begin{aligned} 
\protect\protect\ensuremath{\big\langle 
\protect{\hat{\mathcal Y}}(t_2)\protect{\hat{\mathcal A}}(t_1)
\big\rangle} = \int \frac{d^2 \alpha_2 d^2 \alpha_1 d^2 \alpha _0}{\pi^3 }Y(\alpha _2) 
\protect\protect\ensuremath{\bigg[
\protect\ensuremath{\Big(
\alpha_1 - \frac{1}{2} \frac{\partial }{\partial \alpha_1 ^*} \Big)}
U(\alpha _2,t_2,\alpha _1,t_1) 
\bigg]} 
U(\alpha _1,t_1,\alpha _0,t_0)\rho (\alpha _0,t_0) . 
\end{aligned}}}
% \nonumber 
\label{eq:YA} 
\end{align}}%
%+++++++++++++++++++++++++++++++++++++++++++++
Integration by parts was used to move the derivative to $U(\alpha _2,t_2,\alpha _1,t_1)$; square brackets emphasize that the differentiation does not apply to $U(\alpha _1,t_1,\alpha _0,t_0)$. Similar considerations yield 
%=============================================
\protect{\begin{align}{{
 \begin{aligned} 
\protect\protect\ensuremath{\big\langle 
\protect{\hat{\mathcal A}}(t_1)\protect{\hat{\mathcal Y}}(t_2)
\big\rangle} = \int \frac{d^2 \alpha_2 d^2 \alpha_1 d^2 \alpha _0}{\pi^3 }Y(\alpha _2) 
%\\ \times
\protect\protect\ensuremath{\bigg[
\protect\ensuremath{\Big(
\alpha_1 + \frac{1}{2} \frac{\partial }{\partial \alpha_1 ^*} \Big)}
U(\alpha _2,t_2,\alpha _1,t_1) 
\bigg]} %\\ \times
U(\alpha _1,t_1,\alpha _0,t_0)\rho (\alpha _0,t_0) , 
\end{aligned}}}
% \nonumber 
\label{eq:AY} 
\end{align}}%
%+++++++++++++++++++++++++++++++++++++++++++++
%=============================================
\protect{\begin{align}{{
 \begin{aligned} 
\protect\protect\ensuremath{\big\langle 
\protect{\hat{\mathcal A}}^{\dag}(t_1)\protect{\hat{\mathcal Y}}(t_2)
\big\rangle} = \int \frac{d^2 \alpha_2 d^2 \alpha_1 d^2 \alpha _0}{\pi^3 }Y(\alpha _2) 
%\\ \times
\protect\protect\ensuremath{\bigg[
\protect\ensuremath{\Big(
\alpha_1^* - \frac{1}{2} \frac{\partial }{\partial \alpha_1 } \Big)}
U(\alpha _2,t_2,\alpha _1,t_1) 
\bigg]} %\\ \times
U(\alpha _1,t_1,\alpha _0,t_0)\rho (\alpha _0,t_0) , 
\end{aligned}}}
% \nonumber 
\label{eq:AdY} 
\end{align}}%
%+++++++++++++++++++++++++++++++++++++++++++++
%=============================================
\protect{\begin{align}{{
 \begin{aligned} 
\protect\protect\ensuremath{\big\langle 
\protect{\hat{\mathcal Y}}(t_2)\protect{\hat{\mathcal A}}^{\dag}(t_1)
\big\rangle} = \int \frac{d^2 \alpha_2 d^2 \alpha_1 d^2 \alpha _0}{\pi^3 }Y(\alpha _2) 
%\\ \times
\protect\protect\ensuremath{\bigg[
\protect\ensuremath{\Big(
\alpha_1^* + \frac{1}{2} \frac{\partial }{\partial \alpha_1} \Big)}
U(\alpha _2,t_2,\alpha _1,t_1) 
\bigg]} %\\ \times
U(\alpha _1,t_1,\alpha _0,t_0)\rho (\alpha _0,t_0) . 
\end{aligned}}}
% \nonumber 
\label{eq:YAd} 
\end{align}}%
%+++++++++++++++++++++++++++++++++++++++++++++
We remind the reader that eqs.\ (\protect\ref{eq:YA})--(\ref{eq:YAd}) hold if $t_0<t_1<t_2$. Note that the latest operator in them remains arbitrary. 

%******************************************* 
\subsection{Commuting Heisenberg operators as the quantum response problem}
%******************************************* 
Relations (\ref{eq:AY})--(\ref{eq:YAd}) are exact and not associated with the path-integral representation of the phase-space amplitude. However their most natural interpretation is in terms of path-integral averages. Adding (\ref{eq:AY}) and (\ref{eq:YA}) together we find 
%=============================================
\protect{\begin{align}{{
 \begin{aligned} 
\frac{1}{2} \protect\protect\ensuremath{\big\langle 
\protect{\hat{\mathcal Y}}(t_2)\protect{\hat{\mathcal A}}(t_1) +
\protect{\hat{\mathcal A}}(t_1)\protect{\hat{\mathcal Y}}(t_2)
\big\rangle} = \int \frac{d^2 \alpha_2 d^2 \alpha_1 d^2 \alpha _0}{\pi^3 }Y(\alpha _2) \alpha_1 
U(\alpha _2,t_2,\alpha _1,t_1) 
%\\ \times
U(\alpha _1,t_1,\alpha _0,t_0)\rho (\alpha _0,t_0). 
\end{aligned}}}
% \nonumber % \eqlabel{} 
\end{align}}%
%+++++++++++++++++++++++++++++++++++++++++++++
This is nothing but the path-integral average of $Y(\alpha(t _2)) \alpha(t_1)$ \cite{But3}, where $\alpha (t)$ is an individual path. Hence 
%=============================================
\protect{\begin{gather}{{
 \begin{gathered} 
\frac{1}{2} \protect\protect\ensuremath{\big\langle 
\protect{\hat{\mathcal Y}}(t_2)\protect{\hat{\mathcal A}}(t_1) +
\protect{\hat{\mathcal A}}(t_1)\protect{\hat{\mathcal Y}}(t_2)
\big\rangle} = \overline{\hspace{0.1ex}Y(\alpha(t _2)) \alpha(t_1)\hspace{0.1ex}} , \\ 
\frac{1}{2} \protect\protect\ensuremath{\big\langle 
\protect{\hat{\mathcal Y}}(t_2)\protect{\hat{\mathcal A}}^{\dag}(t_1) +
\protect{\hat{\mathcal A}}^{\dag}(t_1)\protect{\hat{\mathcal Y}}(t_2)
\big\rangle} = \overline{\hspace{0.1ex}Y(\alpha(t _2)) \alpha^*(t_1)\hspace{0.1ex}} . 
\end{gathered}}}
% \nonumber % \eqlabel{} 
\end{gather}}%
%+++++++++++++++++++++++++++++++++++++++++++++
where the bar denotes the (quasi-statistical) averaging over the paths. 
The second line here follows by adding together (\ref{eq:AdY}) and (\ref{eq:YAd}).
These relations suggest that the path-integral averages in the Wigner representation correspond to quantum averages of symmetrised products of Heisenberg field operators. 
Natural as it looks, this result in fact holds only for two-time averages. For three and more times, a new type of operator ordering is found, see section \ref{ch:MultT}. 

It is easy to see that derivatives in eqs.\ (\protect\ref{eq:YA})--(\ref{eq:YAd}) correspond to commutators of the Heisenberg operators. For example, subtracting (\ref{eq:AdY}) from (\ref{eq:YAd}) yields 
%=============================================
\protect{\begin{align}{{
 \begin{aligned} 
\protect\protect\ensuremath{\big\langle \protect\protect\ensuremath{\big[
\protect{\hat{\mathcal Y}}(t_2),\protect{\hat{\mathcal A}}^{\dag}(t_1)
\big]} 
\big\rangle} = \int \frac{d^2 \alpha_2 d^2 \alpha_1 d^2 \alpha _0}{\pi^3 }Y(\alpha _2) 
%\\ \times
\frac{\partial U(\alpha _2,t_2,\alpha _1,t_1)}{\partial \alpha_1} 
%\\ \times
U(\alpha _1,t_1,\alpha _0,t_0)\rho (\alpha _0,t_0) , 
\end{aligned}}}
% \nonumber 
\label{eq:Comm} 
\end{align}}%
%+++++++++++++++++++++++++++++++++++++++++++++
\end{widetext}%
As was suggested by Polkovnikov \cite{Polkan}, quantities like (\ref{eq:Comm}) may be calculated numerically within a path-integral approach by introducing ``quantum jumps'' in the trajectories. In practice, this means splitting a trajectory at $t=t_1$ by introducing small complex shifts $\alpha (t_1)\to \alpha (t_1)+\delta \alpha $, in order to calculate the derivative numerically. For a complex derivative, two independent shifts are needed, hence for $t>t_1$ three trajectories must be run in parallel. The numerical cost of such ``quantum jumps'' is obviously not prohibitive. 

A similar method was employed in \cite{CommP} to calculate symmetrised averages of Heisenberg operators using the so-called positive-P representation. Path-integral averages within the latter correspond to time-normally ordered operator averages. The commutators in \cite{CommP} were expressed by, firstly, using Kubo's famous formula for the linear response function \cite{Kubo} to associate the commutator with response of the quantum system, and, secondly, connecting the latter to a response problem in phase space. It is easy to see that such response interpretation also holds for the derivatives in eqs.\ (\protect\ref{eq:YA})--(\ref{eq:YAd}). Indeed, assume that the Hamiltonian of the system has been complemented by external c-number sources, 
%=============================================
\protect{\begin{align}{{
 \begin{aligned} 
\hat H(t) \to \Hat H(t) - s(t)\hat a^{\dag} - s^*(t) \hat a . 
\end{aligned}}}
% \nonumber % \eqlabel{} 
\end{align}}%
%+++++++++++++++++++++++++++++++++++++++++++++
In phase-space terms, this only modifies the regular evolution, 
%=============================================
\protect{\begin{gather}{{
 \begin{gathered} 
f(\alpha ,t) \to f(\alpha ,t) - s(t), \\ f^*(\alpha ,t) \to f^*(\alpha ,t) - s^*(t), 
\end{gathered}}}
% \nonumber % \eqlabel{} 
\end{gather}}%
%+++++++++++++++++++++++++++++++++++++++++++++
cf.\ eqs.\ (\protect\ref{eq:UInf})--(\ref{eq:UInf1}). 
Shifting a trajectory by $\delta \alpha $ at $t=t_1$ requires an instantaneous source 
%=============================================
\protect{\begin{align}{{
 \begin{aligned} 
s(t) = -i\hbar\, \delta \alpha\, \delta (t-t_1) . 
\end{aligned}}}
% \nonumber % \eqlabel{} 
\end{align}}%
%+++++++++++++++++++++++++++++++++++++++++++++
We thus have a simple correspondence 
between sources and ``quantum jumps:'' (cf.\ also endnote \cite{But3})
%=============================================
\protect{\begin{align}{{
 \begin{aligned} 
\frac{\delta }{\delta s(t_1)} &\Longleftrightarrow \frac{i}{\hbar } \frac{\partial }{\partial \alpha _1}, &
\frac{\delta }{\delta s^*(t_1)} &\Longleftrightarrow - \frac{i}{\hbar } \frac{\partial }{\partial \alpha^* _1} . 
\end{aligned}}}
% \nonumber % \eqlabel{} 
\end{align}}%
%+++++++++++++++++++++++++++++++++++++++++++++ 
The RHS of (\ref{eq:Comm}) is then recognised to be, up to a factor, the linear response function of the quantum system. We can therefore rewrite (\ref{eq:Comm}) as 
%=============================================
\protect{\begin{align}{{
 \begin{aligned} 
\frac{\delta \protect\protect\ensuremath{\big\langle 
\protect{\hat{\mathcal Y}}(t_2)
\big\rangle} }{\delta s(t_1)} \bigg |_{s=0} = \frac{i}{\hbar } 
\protect\protect\ensuremath{\Big\langle \protect\protect\ensuremath{\big[
\protect{\hat{\mathcal Y}}(t_2),\protect{\hat{\mathcal A}}^{\dag}(t_1)
\big]} 
\Big\rangle}, \ \ t_2>t_1.
\end{aligned}}}
% \nonumber 
\label{eq:Kubo} 
\end{align}}%
%+++++++++++++++++++++++++++++++++++++++++++++
This is nothing but Kubo's formula for the linear response function. Hence equation (\ref{eq:Comm}) can be derived starting from Kubo's formula and then using the correspondence between the sources and ``quantum jumps.'' A general discussion of the quantum response problem in the Wigner representation requires advanced formal tools and will be presented elsewhere. 
%******************************************* 
\subsection{Multi-time averages and time-symmetric operator ordering%
\label{ch:MultT}} 
%******************************************* 
Consider the path-integral average ($t_1<t_2<\cdots t_N$)
%=============================================
\protect{\begin{align}{{
 \begin{aligned} 
\overline{\hspace{0.1ex}\alpha(t_1)\alpha(t_2)\cdots\alpha(t_N) \hspace{0.1ex}} \equiv \protect\protect\ensuremath{\big\langle 
{\cal T}_W \protect{\hat{\mathcal A}}(t_1)\protect{\hat{\mathcal A}}(t_2)\cdots\protect{\hat{\mathcal A}}(t_N) 
\big\rangle} , 
\end{aligned}}}
% \nonumber 
\label{eq:TW} 
\end{align}}%
%+++++++++++++++++++++++++++++++++++++++++++++
where the Heisenberg field operators are given by (\ref{eq:HeisA}). The LHS of eq.\ (\protect\ref{eq:TW}) is defined in phase-space terms as 
\begin{widetext} 
%=============================================
\protect{\begin{align}{{
 \begin{aligned} 
\overline{\hspace{0.1ex}\alpha(t_1)\alpha(t_2)\cdots\alpha(t_N)\hspace{0.1ex}} = \int
\frac{d^2\alpha_0 d^2\alpha_1\cdots d^2\alpha_N}{\pi ^{N+1}}
%\\ \times 
\rho (\alpha _0,t_0) 
\prod_{k=1}^N 
\alpha _k U(\alpha _k,t_k,\alpha _{k-1},t_{k-1}) . 
\end{aligned}}}
% \nonumber 
\label{eq:TWI} 
\end{align}}%
%+++++++++++++++++++++++++++++++++++++++++++++
On the RHS of (\ref{eq:TW}) we redefine it in the Hilbert-space terms as an average of a {\em time-symmetric product\/} of the field operators. 

To obtain an independent characterisation of the time-symmetric ordering we use the same trick that lead to eqs.\ (\protect\ref{eq:YA})--(\ref{eq:YAd}), but apply it ``in reverse.'' We start from defining the quantity 
%=============================================
\protect{\begin{align}{{
 \begin{aligned} 
B(\alpha_2,t_2) = 
\int
\frac{ d^2\alpha_3\cdots d^2\alpha_N}{\pi ^{N-2}}
%\\ \times 
\alpha _2 \prod_{k=3}^N 
\alpha _k U(\alpha _k,t_k,\alpha _{k-1},t_{k-1}) . 
\end{aligned}}}
% \nonumber % \eqlabel{} 
\end{align}}%
%+++++++++++++++++++++++++++++++++++++++++++++
The times $t_3,\cdots,t_N$ here should be considered as parameters. Following the pattern of eqs.\ (\protect\ref{eq:BU}) and (\ref{eq:Usf}) we then write 
%=============================================
\protect{\begin{align}{{
 \begin{aligned} 
\int \frac{d^2 \alpha_2}{\pi }B(\alpha_2,t_2)U(\alpha _2,t_2,\alpha _{1},t_{1}) = 
\protect\protect\ensuremath{\big[
\protect{\hat{\mathcal B}}_{t_1}(t_2)
\big]} (\alpha_1) = 
\protect\protect\ensuremath{\big[
\protect{\hat{\mathcal U}}(t_1,t_0)\protect{\hat{\mathcal B}}(t_2)\protect{\hat{\mathcal U}}^{\dag}(t_1,t_0)
\big]} (\alpha_1) . 
\end{aligned}}}
% \nonumber 
\label{eq:Bt2} 
\end{align}}%
%+++++++++++++++++++++++++++++++++++++++++++++
Furthermore, by making use of eq.\ (\protect\ref{eq:Rt1}) and of the phase-space correspondences, we have 
%=============================================
\protect{\begin{align}{{
 \begin{aligned} 
\alpha _1\int \frac{d^2 \alpha _{0}}{\pi }U(\alpha _1,t_1,\alpha _{0},t_{0})
\rho (\alpha _{0},t_{0}) = \alpha _1\rho (\alpha _{1},t_{1})
= \frac{1}{2} 
\protect\protect\ensuremath{\big[
\hat a\hat \rho (t_1)+\hat \rho (t_1)\hat a
\big]} (\alpha _{1})
\end{aligned}}}
% \nonumber 
\label{eq:RecR} 
\end{align}}%
%+++++++++++++++++++++++++++++++++++++++++++++
The remaining integration over $\alpha _{1}$ in (\ref{eq:TWI}) is then performed straghtaway and we find 
%=============================================
\protect{\begin{align}{{
 \begin{aligned} 
\protect\protect\ensuremath{\big\langle 
{\cal T}_W \protect{\hat{\mathcal A}}(t_1)\protect{\hat{\mathcal A}}(t_2)\cdots\protect{\hat{\mathcal A}}(t_N) 
\big\rangle} = \frac{1}{2}\text{Tr}\, \protect{\hat{\mathcal U}}(t_1,t_0)\protect{\hat{\mathcal B}}(t_2)\protect{\hat{\mathcal U}}^{\dag}(t_1,t_0)\protect\protect\ensuremath{\big[
\hat a\hat \rho (t_1)+\hat \rho (t_1)\hat a
\big]} = \frac{1}{2}\protect\protect\ensuremath{\big\langle 
\protect\protect\ensuremath{\big\{
\protect{\hat{\mathcal B}}(t_2),\protect{\hat{\mathcal A}}(t_1)
\big\}} 
\big\rangle} , 
\end{aligned}}}
% \nonumber 
\label{eq:Rec0} 
\end{align}}%
%+++++++++++++++++++++++++++++++++++++++++++++
%\end{widetext}%
where the curly brackets stand for the anticommutator, 
%=============================================
\protect{\begin{align}{{
 \begin{aligned} 
\protect\protect\ensuremath{\big\{
\hat X, \hat Y
\big\}} = \hat X \hat Y + \hat Y \hat X. 
\end{aligned}}}
% \nonumber % \eqlabel{} 
\end{align}}%
%+++++++++++++++++++++++++++++++++++++++++++++
The last equality in (\ref{eq:Rec0}) follows from 
%=============================================
\protect{\begin{align}{{
 \begin{aligned} 
\hat \rho (t_1) = \protect{\hat{\mathcal U}}(t_1,t_0)\hat \rho (t_0)\protect{\hat{\mathcal U}}^{\dag}(t_1,t_0) . 
\end{aligned}}}
% \nonumber % \eqlabel{} 
\end{align}}%
%+++++++++++++++++++++++++++++++++++++++++++++
Finally, the operator $\protect{\hat{\mathcal B}}(t_2)$ is found if we take eq.\ (\protect\ref{eq:Bt2}) at $t_1=t_0$ and compare it with eq.\ (\protect\ref{eq:TWI}). It is then easy to see that 
%=============================================
\protect{\begin{align}{{
 \begin{aligned} 
\protect{\hat{\mathcal B}}(t_2) = {\cal T}_W\protect{\hat{\mathcal A}}(t_2)\cdots\protect{\hat{\mathcal A}}(t_N). 
\end{aligned}}}
% \nonumber 
\label{eq:Bt2N} 
\end{align}}%
%+++++++++++++++++++++++++++++++++++++++++++++
Equation (\ref{eq:Rec0}) is thus a recursive relation for the time-ordered operator products. 

If we replace in (\ref{eq:TW}) $\alpha (t_1)\to \alpha ^*(t_1)$ and 
$\protect{\hat{\mathcal A}}(t_1)\to\protect{\hat{\mathcal A}}^{\dag}(t_1)$, equation (\ref{eq:Rec0}) will also hold with $\protect{\hat{\mathcal A}}(t_1)\to\protect{\hat{\mathcal A}}^{\dag}(t_1)$. Furthermore, the actual nature of the factors $\alpha (t_2)\cdots\alpha (t_N)$ is irrevant. Any subset of them may be complex-conjugated, provided the corresponding operators under the ${\cal T}_W$-ordering are Hermitian-conjugated. As a result, we arrive at the following recursive definition of the time-symmetric ordering:
%=============================================
\protect{\begin{gather}{{
 \begin{gathered} 
{\cal T}_W\hat{\openone} = \hat{\openone}, \\ 
{\cal T}_W\protect{\hat{\mathcal A}}(t) = \protect{\hat{\mathcal A}}(t), \ \ \ 
{\cal T}_W\protect{\hat{\mathcal A}}^{\dag}(t) = \protect{\hat{\mathcal A}}^{\dag}(t), \\ 
{\cal T}_W \protect{\hat{\mathcal A}}(t) \protect{\hat{\mathcal P}}_{>t} = \frac{1}{2} \protect\protect\ensuremath{\big\{
\protect{\hat{\mathcal A}}(t),\protect{\hat{\mathcal P}}_{>t}
\big\}}, \\ 
{\cal T}_W \protect{\hat{\mathcal A}}^{\dag}(t) \protect{\hat{\mathcal P}}_{>t} = \frac{1}{2} \protect\protect\ensuremath{\big\{
\protect{\hat{\mathcal A}}^{\dag}(t),\protect{\hat{\mathcal P}}_{>t} 
\big\}} ,
\end{gathered}}}
% \nonumber % \eqlabel{} 
\end{gather}}%
%+++++++++++++++++++++++++++++++++++++++++++++
where $\protect{\hat{\mathcal P}}_{>t}$ is a time-symmetric product of operators with all time arguments larger than $t$. 

The number of different terms in a time-symmetric product of $N$ factors is $2^{N-1}$. For $N>2$, this is clearly different from $N!$ permutations comprising a symmetrised product of the same factors. The time-symmetric and symmetrised products are thus distinct for all $N>2$. For example, for $N=3$ and $t_1<t_2<t_3$ we have 
%\begin{widetext} 
%=============================================
\protect{\begin{align}{{
 \begin{aligned} 
{\cal T}_W \protect{\hat{\mathcal A}}(t_1)\protect{\hat{\mathcal A}}(t_2)\protect{\hat{\mathcal A}}(t_3) = 
\frac{1}{4} \protect\protect\ensuremath{\big[
\protect{\hat{\mathcal A}}(t_1)\protect{\hat{\mathcal A}}(t_2)\protect{\hat{\mathcal A}}(t_3) + 
\protect{\hat{\mathcal A}}(t_2)\protect{\hat{\mathcal A}}(t_3)\protect{\hat{\mathcal A}}(t_1) + 
\protect{\hat{\mathcal A}}(t_1)\protect{\hat{\mathcal A}}(t_3)\protect{\hat{\mathcal A}}(t_2) + 
\protect{\hat{\mathcal A}}(t_3)\protect{\hat{\mathcal A}}(t_2)\protect{\hat{\mathcal A}}(t_1) 
\big]} . 
\end{aligned}}}
% \nonumber % \eqlabel{} 
\end{align}}%
%+++++++++++++++++++++++++++++++++++++++++++++
The symmetrised product of the same factors should also include $\protect{\hat{\mathcal A}}(t_2)\protect{\hat{\mathcal A}}(t_1)\protect{\hat{\mathcal A}}(t_3)$ and $\protect{\hat{\mathcal A}}(t_3)\protect{\hat{\mathcal A}}(t_1)\protect{\hat{\mathcal A}}(t_2)$. 

The most important properties of the time-symmetric products are the two: these products are continuous at coinciding time arguments, and for free-field operators they turn into the conventional symmetric (Weyl) ordered products. A proof of these properties, as well as a more detailed discussion of the time symmetric ordering, including its relation to Schwinger's closed-time-loop formalism, will be presented elsewhere. 

%******************************************* 
\section{Multitime representations based on other orderings}\label{ch:S} 
%******************************************* 
\subsection{Definitions} 
%******************************************* 
Non-symmetrically ordered representations are introduced by replacing eq.\ (\protect\ref{eq:EqW}) by 
%=============================================
\protect{\begin{align}{{
 \begin{aligned} 
A_s(\alpha) = \protect\protect\ensuremath{\big[
\hat A
\big]}_s (\alpha) = \int \frac{d^2 \beta }{\pi }\text{e}^{
\beta\alpha^* - \beta^*\alpha + s|\beta |^2/2 }
%\\ \times 
\text{Tr} \hat A \hat D^{\dag}(\beta ) 
= \exp\protect\ensuremath{\Big(
{ 
-\frac{s}{2} \,{\frac{\partial^2 }
{\partial\alpha \partial\alpha^* }}
} \Big)}
A(\alpha). 
\end{aligned}}}
% \nonumber 
\label{eq:EqS} 
\end{align}}%
%+++++++++++++++++++++++++++++++++++++++++++++
Equation (\ref{eq:TrAB}) then becomes 
%=============================================
\protect{\begin{align}{{
 \begin{aligned} 
\text{Tr} \hat A \hat B 
= \int \frac{d^2 \alpha}{\pi } \protect\protect\ensuremath{\bigg[
\exp\protect\ensuremath{\Big(
{ 
-\frac{s}{2} \,{\frac{\partial^2 }
{\partial\alpha \partial\alpha^* }}
} \Big)}A(\alpha)
\bigg]} \protect\protect\ensuremath{\bigg[
\exp\protect\ensuremath{\Big(
{ 
\frac{s}{2} \,{\frac{\partial^2 }
{\partial\alpha \partial\alpha^* }}
} \Big)}B(\alpha)
\bigg]} 
= \int \frac{d^2 \alpha}{\pi } A_s(\alpha)B_{-s}(\alpha) . 
\end{aligned}}}
% \nonumber 
\label{eq:TrABs} 
\end{align}}%
%+++++++++++++++++++++++++++++++++++++++++++++
\end{widetext}%
The normal, Weyl and anti-normal representations of the $\rho $-matrix, 
corresponding to $s=-1,0,1$, are commonly known as $Q$, $W$ and $P$ functions, respectively. They are also called quasi-distributions, because, for example, the $P$-function gives a classically-looking expression for averages of normally-ordered operator expressions, 
%=============================================
\protect{\begin{align}{{
 \begin{aligned} 
\text{Tr} \hat \rho \,{\mbox{\rm\boldmath$:$}} F(\hat a,\hat a^{\dag}){\mbox{\rm\boldmath$:$}} 
= \int \frac{d^2\alpha}{\pi } P(\alpha)
F(\alpha,\alpha^*), 
\end{aligned}}}
% \nonumber % \eqlabel{} 
\end{align}}%
%+++++++++++++++++++++++++++++++++++++++++++++
and similarly for the $Q$-function with the antinormal operator ordering. Thus normal representation of field operators is associated with antinormally ordered $\rho $-matrix, and {\em vice versa\/}. 
To avoid lengthy repeated specifications, we introduce the term $s${\em -representation\/} for a phase-space picture where the field operators are $s$-ordered and the $\rho $-matrix is $(-s)$-ordered. 
This agrees fully with the use of this term in quantum optics: normal representation employs $P$-function while antinormal representation employs $Q$-function, and so on. 
%******************************************* 
\subsection{Phase-space transition amplitude} 
%******************************************* 
Manipulating (\ref{eq:BAv}) similar to (\ref{eq:TrABs}) yields in the $s$-representation 
%=============================================
\protect{\begin{align}{{
 \begin{aligned} 
\protect\protect\ensuremath{\big\langle 
\protect{\hat{\mathcal B}}(t)
\big\rangle} = \int \frac{d^2 \alpha _0d^2 \alpha }{\pi^2} 
B_s(\alpha ,t)U_s(\alpha ,t,\alpha _0,t_0)\rho_{-s} (\alpha,t_0), 
\end{aligned}}}
% \nonumber 
\label{eq:BAvS} 
\end{align}}%
%+++++++++++++++++++++++++++++++++++++++++++++
where $B_s(\alpha ,t) = \protect\protect\ensuremath{\big[
\hat B(t)
\big]}_s (\alpha)$ and 
\begin{widetext} 
%=============================================
{\begin{multline} 
U_s(\alpha ,t,\alpha _0,t_0) = 
\exp\protect\protect\ensuremath{\Big[
\frac{s}{2} \,\protect\ensuremath{\Big(
{ 
{\frac{\partial^2 }
{\partial\alpha \partial\alpha^* }}
%} }\exp\Rbracket{\Big}{{ 
-{\frac{\partial^2 }
{\partial\alpha_0 \partial\alpha_0^* }}
} \Big)}\Big]}U(\alpha ,t,\alpha _0,t_0) \\ = 
\int \frac{d^2 \beta _0d^2 \beta }{\pi^2}\text{e}^{
\alpha\beta^* - \alpha^*\beta + \alpha_0\beta_0^* - \alpha_0^*\beta_0 
+ s(|\beta_0 |^2 - |\beta |^2)/2}
%\\ \times 
\,\text{Tr}\hat D^{\dag}(\beta_0 )\protect{\hat{\mathcal U}}^{\dag}(t,t_0)\hat D^{\dag}(\beta )\protect{\hat{\mathcal U}}(t,t_0)
, 
% \nonumber 
\label{eq:UDefS} 
\end{multline}}%
%+++++++++++++++++++++++++++++++++++++++++++++
%\end{widetext}%
is the phase-space transition amplitude in the $s$-representation. It is easy to see that the properties (\ref{eq:Dlt})--(\ref{eq:UU}) also hold for $U_s$, while eqs.\ (\protect\ref{eq:Rt1})--(\ref{eq:Usf}) become 
%=============================================
\protect{\begin{align}{{
 \begin{aligned} 
\int \frac{d^2 \alpha_0}{\pi }U_s(\alpha_1 ,t_1,\alpha _0,t_0)
\rho_{-s} (\alpha _0,t_0) 
%\\ 
= 
\rho_{-s} (\alpha_1 ,t_1) = 
\protect\protect\ensuremath{\big[
\hat \rho (t_1)
\big]}_{-s} (\alpha_1)
\end{aligned}}}
% \nonumber 
\label{eq:Rt1S} 
\end{align}}%
%+++++++++++++++++++++++++++++++++++++++++++++
and 
%=============================================
\protect{\begin{align}{{
 \begin{aligned} 
\int \frac{d^2 \alpha}{\pi }B_s(\alpha,t)U_s(\alpha ,t,\alpha _1,t_1) = 
\protect\protect\ensuremath{\big[
\protect{\hat{\mathcal B}}_{t_1}(t)
\big]}_s (\alpha_1) 
%\\ 
= \protect\protect\ensuremath{\big[
\protect{\hat{\mathcal U}}(t_1,t_0)\protect{\hat{\mathcal B}}(t)\protect{\hat{\mathcal U}}^{\dag}(t_1,t_0)
\big]}_s (\alpha_1). 
\end{aligned}}}
% \nonumber 
\label{eq:UsfS} 
\end{align}}%
%+++++++++++++++++++++++++++++++++++++++++++++
%\end{widetext}%
%******************************************* 
\subsection{Path-integral representation of the amplitude} 
%******************************************* 
While the general structural properties of the path integral in phase space do not change with the ordering, the path integral itself is certainly ordering-specific. Consider the infinitesimal amplitude $U(\alpha,t+\Delta t,\alpha_0,t)$. Combining eqs.\ (\protect\ref{eq:UInf}) and (\ref{eq:UDefS}) we have 
%\begin{widetext} 
%=============================================
{\begin{multline} 
U_s(\alpha ,t+\Delta t,\alpha _0,t) = 
\exp\protect\protect\ensuremath{\bigg[
\frac{s}{2} \,\protect\ensuremath{\Big(
{ 
{\frac{\partial^2 }
{\partial\alpha \partial\alpha^* }}
-{\frac{\partial^2 }
{\partial\alpha_0 \partial\alpha_0^* }}
} \Big)}\bigg]} 
%\\ \times 
\int \frac{d^2 \sigma }{\pi }
\exp\protect\protect\ensuremath{\big[
(\alpha -\alpha _0)\sigma ^* - (\alpha -\alpha _0)^*\sigma
\big]} 
\\ \times 
\protect\protect\ensuremath{\bigg\{
1 - \frac{i\Delta t}{\hbar } \protect\protect\ensuremath{\Big[
H(\alpha_0 - \sigma /2,t) - H(\alpha_0 + \sigma /2,t)
\Big]}
\bigg\}} . 
% \nonumber 
\label{eq:UInfS} 
\end{multline}}%
%+++++++++++++++++++++++++++++++++++++++++++++
A straightforward calculation in Appendix \ref{app:InfUS} yields
%=============================================
{\begin{multline} 
U_s(\alpha ,t+\Delta t,\alpha _0,t) = 
\int \frac{d^2 \sigma }{\pi }\exp
\protect\protect\ensuremath{\bigg\{
(\alpha -\alpha _0)\sigma ^* - (\alpha -\alpha _0)^*\sigma
\\ - \frac{i\Delta t}{\hbar } \protect\protect\ensuremath{\bigg[
H_s\protect\ensuremath{\Big(
\alpha_0 - \frac{(1+s)\sigma }{2},
\alpha^*_0 - \frac{(1-s)\sigma^*}{2} ,t
 \Big)} 
- 
H_s\protect\ensuremath{\Big(
\alpha_0 + \frac{(1-s)\sigma}{2},
\alpha^*_0 + \frac{(1+s)\sigma^*}{2},t
 \Big)} 
\bigg]}
\bigg\}} . 
% \nonumber 
\label{eq:UInfS1} 
\end{multline}}%
%+++++++++++++++++++++++++++++++++++++++++++++
We had to amend our notation, 
%=============================================
\protect{\begin{align}{{
 \begin{aligned} 
H_s(\alpha,t)\to H_s(\alpha,\alpha^*,t), 
\end{aligned}}}
% \nonumber % \eqlabel{} 
\end{align}}%
%+++++++++++++++++++++++++++++++++++++++++++++
to accommodate for asymmetry between $\alpha$ and $\alpha^*$ introduced by 
a non-symmetric ordering. 
Separating the deterministic part of the motion in phase space we get in place of eq.\ (\protect\ref{eq:UInf1})
%=============================================
\protect{\begin{align}{{
 \begin{aligned} 
U_s(\alpha ,t+\Delta t,\alpha _0,t) = \int \frac{d^2 \sigma }{\pi }\exp\protect\protect\ensuremath{\bigg\{
(\alpha -\alpha _0 + if_s\Delta t /\hbar )\sigma ^* - (\alpha -\alpha _0 + if_s\Delta t /\hbar )^*\sigma + \frac{i\Delta t}{\hbar }h_s(\alpha_0,\sigma,t) 
\bigg\}} , 
\end{aligned}}}
% \nonumber 
\label{eq:UInf1S} 
\end{align}}%
%+++++++++++++++++++++++++++++++++++++++++++++
where 
%=============================================
\protect{\begin{align}{{
 \begin{aligned} 
f_s(\alpha_0,t) &= \frac{\partial H_s(\alpha_0 ,t)}{\partial \alpha_0^*}, & 
f_s^*(\alpha_0,t) &= 
\frac{\partial H_s(\alpha_0 ,t)}{\partial \alpha_0}, 
\end{aligned}}}
% \nonumber % \eqlabel{} 
\end{align}}%
%+++++++++++++++++++++++++++++++++++++++++++++
and 
%=============================================
{\begin{multline} 
h_s(\alpha_0,\sigma,t) = 
H_s\protect\ensuremath{\Big(
\alpha_0 + \frac{(1-s)\sigma}{2},
\alpha^*_0 + \frac{(1+s)\sigma^*}{2},t
 \Big)} 
\\ - 
H_s\protect\ensuremath{\Big(
\alpha_0 - \frac{(1+s)\sigma }{2},
\alpha^*_0 - \frac{(1-s)\sigma^*}{2} ,t
 \Big)} 
- \sigma f_s^*(\alpha_0,t) - \sigma ^* f_s(\alpha_0,t). 
% \nonumber % \eqlabel{} 
\end{multline}}%
%+++++++++++++++++++++++++++++++++++++++++++++
%\end{widetext}%
Expanding this into power series we get for the practically important cases of the normal, antinormal, and Wigner representations
%=============================================
\protect{\begin{align}{{
 \begin{aligned} 
h_1(\alpha_0,\sigma,t) 
&= \sum _{k\geq 2}\frac{
f_1^{(0,k)}\sigma ^{*k} - f_1^{(k,0)}(-\sigma )^{k}
}{k!} ,
&\textrm{\ ($P$)} \\ 
h_{-1}(\alpha_0,\sigma,t) 
&= \sum _{k\geq 2}\frac{
f_{-1}^{(k,0)}\sigma ^{k} - f_{-1}^{(0,k)}(-\sigma^* )^{k}
}{k!} ,
&\textrm{\ ($Q$)} \\ 
h_{0}(\alpha_0,\sigma,t) 
&= \sum _{
%\scriptsize\begin{array}{c} 
k+m\geq 3,\textrm{odd}
%\end{array}%
}\frac{
f_0^{(k,m)}\sigma ^{k}\sigma ^{*m}
}{2^{k+m-1}k!m!} ,
&\textrm{\ ($W$)} 
\end{aligned}}}
% \nonumber % \eqlabel{} 
\end{align}}%
%+++++++++++++++++++++++++++++++++++++++++++++
where 
%=============================================
\protect{\begin{gather}{{
 \begin{gathered} 
f_s^{(k,m)} = f_s^{(k,m)}(\alpha_0,t) = 
\frac{\partial^{k+m} H_s(\alpha_0 ,t)}
{\partial\alpha_0^k \partial\alpha_0^{*m} }, \\ 
\protect\protect\ensuremath{\big[
f_s^{(k,m)}(\alpha_0,t)
\big]}^* = f_s^{(m,k)}(\alpha_0,t). 
\end{gathered}}}
% \nonumber % \eqlabel{} 
\end{gather}}%
%+++++++++++++++++++++++++++++++++++++++++++++

Unlike in the symmetric case, for an arbitrary $s$ the noise contribution starts from the second-order noise. Interestingly, the deterministic path is also ordering-specific. The classical motion is only recovered in the limit $\hbar \to 0$, when {\em both\/} the noise contribution, and the reordering correction in which $H_s(\alpha_0 ,t)$ differ for different $s$, are negligible. 
%******************************************* 
\subsection{Quantum averages} 
%******************************************* 
In the $s$-representation the phase-space correspondences read 
%=============================================
\protect{\begin{gather}{{
 \begin{gathered} 
\hat \rho \hat a \Longleftrightarrow \protect\ensuremath{\Big(
\alpha - \frac{s+1}{2} \frac{\partial }{\partial \alpha ^*} \Big)} 
\rho_{-s}(\alpha), 
\\ 
\hat a \hat \rho \Longleftrightarrow \protect\ensuremath{\Big(
\alpha - \frac{s-1}{2} \frac{\partial }{\partial \alpha ^*} \Big)} 
\rho_{-s}(\alpha), 
\\ 
\hat \rho \hat a^{\dag} \Longleftrightarrow \protect\ensuremath{\Big(
\alpha ^* - \frac{s-1}{2} \frac{\partial }{\partial \alpha} \Big)} 
\rho_{-s}(\alpha), 
\\ 
\hat a^{\dag} \hat \rho \Longleftrightarrow \protect\ensuremath{\Big(
\alpha ^* - \frac{s+1}{2} \frac{\partial }{\partial \alpha} \Big)} 
\rho_{-s}(\alpha). 
\end{gathered}}}
% \nonumber 
\label{eq:CorrS} 
\end{gather}}%
%+++++++++++++++++++++++++++++++++++++++++++++
These formulae may be derived from the definition (\ref{eq:EqS}). In fact, the coefficients at the derivatives may be deduced from eqs.\ (\protect\ref{eq:Corr}) if noticing that, for $s=1$ ($s=-1$), the $\rho $-matrix is antinormally (normally) ordered, hence correspondences for $\hat a \hat \rho $ and $\hat \rho \hat a ^{\dag}$ ($\hat \rho \hat a$ and $\hat a ^{\dag}\hat \rho $) should be without derivatives. Furthermore, the way eqs.\ (\protect\ref{eq:CorrS}) are written emphasizes that the ``nonsymmetric correction'' in these always looks as $-s/2$ times a derivative. The route from the standard correspondences to eqs.\ (\protect\ref{eq:YA})--(\ref{eq:YAd}) involves one integration by parts, hence all ``nonsymmetric corrections'' to (\ref{eq:YA})--(\ref{eq:YAd}) must look as $s/2$ times a derivative. This allows us immediately to write 
%\begin{widetext} 
%=============================================
\protect{\begin{align}{{
 \begin{aligned} 
\protect\protect\ensuremath{\big\langle 
\protect{\hat{\mathcal Y}}(t_2)\protect{\hat{\mathcal A}}(t_1)
\big\rangle} = \int \frac{d^2 \alpha_2 d^2 \alpha_1 d^2 \alpha _0}{\pi^3 }Y_s(\alpha _2) 
\protect\protect\ensuremath{\bigg[
\protect\ensuremath{\Big(
\alpha_1 + \frac{s-1}{2} \frac{\partial }{\partial \alpha_1 ^*} \Big)}
U_s(\alpha _2,t_2,\alpha _1,t_1) 
\bigg]} %\\ \times
U_s(\alpha _1,t_1,\alpha _0,t_0)\rho_{-s}(\alpha _0,t_0) . 
\end{aligned}}}
% \nonumber 
\label{eq:YAS} 
\end{align}}%
%+++++++++++++++++++++++++++++++++++++++++++++
%=============================================
\protect{\begin{align}{{
 \begin{aligned} 
\protect\protect\ensuremath{\big\langle 
\protect{\hat{\mathcal A}}(t_1)\protect{\hat{\mathcal Y}}(t_2)
\big\rangle} = \int \frac{d^2 \alpha_2 d^2 \alpha_1 d^2 \alpha _0}{\pi^3 }Y_s(\alpha _2) 
\protect\protect\ensuremath{\bigg[
\protect\ensuremath{\Big(
\alpha_1 + \frac{s+1}{2} \frac{\partial }{\partial \alpha_1 ^*} \Big)}
U_s(\alpha _2,t_2,\alpha _1,t_1) 
\bigg]} %\\ \times
U_s(\alpha _1,t_1,\alpha _0,t_0)\rho_{-s}(\alpha _0,t_0) , 
\end{aligned}}}
% \nonumber 
\label{eq:AYS} 
\end{align}}%
%+++++++++++++++++++++++++++++++++++++++++++++
%=============================================
\protect{\begin{align}{{
 \begin{aligned} 
\protect\protect\ensuremath{\big\langle 
\protect{\hat{\mathcal A}}^{\dag}(t_1)\protect{\hat{\mathcal Y}}(t_2)
\big\rangle} = \int \frac{d^2 \alpha_2 d^2 \alpha_1 d^2 \alpha _0}{\pi^3 }Y_s(\alpha _2) 
\protect\protect\ensuremath{\bigg[
\protect\ensuremath{\Big(
\alpha_1^* + \frac{s-1}{2} \frac{\partial }{\partial \alpha_1 } \Big)}
U_s(\alpha _2,t_2,\alpha _1,t_1) 
\bigg]} %\\ \times
U_s(\alpha _1,t_1,\alpha _0,t_0)\rho_{-s}(\alpha _0,t_0) , 
\end{aligned}}}
% \nonumber 
\label{eq:AdYS} 
\end{align}}%
%+++++++++++++++++++++++++++++++++++++++++++++
%=============================================
\protect{\begin{align}{{
 \begin{aligned} 
\protect\protect\ensuremath{\big\langle 
\protect{\hat{\mathcal Y}}(t_2)\protect{\hat{\mathcal A}}^{\dag}(t_1)
\big\rangle} = \int \frac{d^2 \alpha_2 d^2 \alpha_1 d^2 \alpha _0}{\pi^3 }Y_s(\alpha _2) 
\protect\protect\ensuremath{\bigg[
\protect\ensuremath{\Big(
\alpha_1^* + \frac{s+1}{2} \frac{\partial }{\partial \alpha_1} \Big)}
U_s(\alpha _2,t_2,\alpha _1,t_1) 
\bigg]} %\\ \times
U_s(\alpha _1,t_1,\alpha _0,t_0)\rho_{-s}(\alpha _0,t_0) . 
\end{aligned}}}
% \nonumber 
\label{eq:YAdS} 
\end{align}}%
%+++++++++++++++++++++++++++++++++++++++++++++
%******************************************* 
\subsection{``Quantum jumps'', response, and the time-$s$-ordering of Heisenberg operators} 
%******************************************* 
The property that adding external sources to the Hamiltonian is equivalent to adding the same sources to the equation for the deterministic path in phase-space does not depend on the operator ordering, hence eqs.\ (\protect\ref{eq:Comm})--(\ref{eq:Kubo}) are valid for any $s$. 
This ``invariance of the response viewpoint'' holds the fact notwithstanding, that the path-integral averages themselves correspond to different types of quantum averages for different values of $s$. On top of purely structural properties of the path integral, the reasoning in section \ref{ch:MultT} 
is based on the relations 
%=============================================
\protect{\begin{align}{{
 \begin{aligned} 
&\frac{1}{2} \protect\protect\ensuremath{\big[
\hat a \hat \rho + \hat \rho \hat a 
\big]} (\alpha) = \alpha \rho(\alpha), & 
\frac{1}{2} \protect\protect\ensuremath{\big[
\hat a^{\dag} \hat \rho + \hat \rho \hat a^{\dag} 
\big]} (\alpha) = \alpha^* \rho(\alpha), 
\end{aligned}}}
% \nonumber 
\label{eq:Symm} 
\end{align}}%
%+++++++++++++++++++++++++++++++++++++++++++++
characteristic of the Weyl ordering. The corresponding relations in the $s$-representation read 
%=============================================
\protect{\begin{align}{{
 \begin{aligned} 
& \protect\protect\ensuremath{\Big[
\frac{1+s}{2}\hat a \hat \rho + \frac{1-s}{2}\hat \rho \hat a
\Big]}_{-s}(\alpha) = \alpha \rho_{-s}(\alpha), & 
\protect\protect\ensuremath{\Big[
\frac{1-s}{2}\hat a^{\dag} \hat \rho + \frac{1+s}{2}\hat \rho \hat a^{\dag}
\Big]}_{-s}(\alpha) = \alpha^* \rho_{-s}(\alpha) . 
\end{aligned}}}
% \nonumber 
\label{eq:SymmS} 
\end{align}}%
%+++++++++++++++++++++++++++++++++++++++++++++
We now define the {\em time-$s$-ordered\/} product of the Heisenberg field operators by generaling eq.\ (\protect\ref{eq:TW}) to the case of $s$-representation, 
%=============================================
\protect{\begin{align}{{
 \begin{aligned} 
\overline{\hspace{0.1ex}\alpha(t_1)\alpha(t_2)\cdots\alpha(t_N) \hspace{0.1ex}}^{(s)} 
\equiv \protect\protect\ensuremath{\big\langle 
{\cal T}_s \protect{\hat{\mathcal A}}(t_1)\protect{\hat{\mathcal A}}(t_2)\cdots\protect{\hat{\mathcal A}}(t_N) 
\big\rangle} . 
\end{aligned}}}
% \nonumber 
\label{eq:TS} 
\end{align}}%
%+++++++++++++++++++++++++++++++++++++++++++++
The path-integral average on the LHS here is given by eq.\ (\protect\ref{eq:TWI}) with $U\to U_s$ and $\rho \to \rho_{-s}$; full meaning to it is then assigned by eq.\ (\protect\ref{eq:UInfS1}). The first relation in \ref{ch:MultT} that has to be modified beyond these trivial replacements is eq.\ (\protect\ref{eq:RecR}): the last equality in it must follow (\ref{eq:SymmS}) instead of (\ref{eq:Symm}). As a result the recursion relation (\ref{eq:Rec0}) changes to 
%=============================================
\protect{\begin{align}{{
 \begin{aligned} 
\protect\protect\ensuremath{\big\langle 
{\cal T}_s \protect{\hat{\mathcal A}}(t_1)\protect{\hat{\mathcal A}}(t_2)\cdots\protect{\hat{\mathcal A}}(t_N) 
\big\rangle} = \protect\protect\ensuremath{\bigg\langle 
\frac{1+s}{2}\protect{\hat{\mathcal B}}^{(s)}(t_2)\protect{\hat{\mathcal A}}(t_1) + 
\frac{1-s}{2}\protect{\hat{\mathcal A}}(t_1)\protect{\hat{\mathcal B}}^{(s)}(t_2)
\bigg\rangle} , 
\end{aligned}}}
% \nonumber 
\label{eq:RecS0} 
\end{align}}%
%+++++++++++++++++++++++++++++++++++++++++++++
\end{widetext}%
with 
%=============================================
\protect{\begin{align}{{
 \begin{aligned} 
\protect{\hat{\mathcal B}}^{(s)}(t_2) = {\cal T}_s\protect{\hat{\mathcal A}}(t_2)\cdots\protect{\hat{\mathcal A}}(t_N). 
\end{aligned}}}
% \nonumber 
\label{eq:Bt2NS} 
\end{align}}%
%+++++++++++++++++++++++++++++++++++++++++++++
Similar reasoning applies with $\protect{\hat{\mathcal A}}(t_1)\to\protect{\hat{\mathcal A}}^{\dag}(t_1)$, except that with this replacement one should also swap the weight factors in (\ref{eq:RecS0}). We thus arrive at the recursive definition of the time-$s$-ordered product: 
%=============================================
\protect{\begin{gather}{{
 \begin{gathered} 
{\cal T}_s\hat{\openone} = \hat{\openone}, \\ 
{\cal T}_s\protect{\hat{\mathcal A}}(t) = \protect{\hat{\mathcal A}}(t), \ \ \ 
{\cal T}_s\protect{\hat{\mathcal A}}^{\dag}(t) = \protect{\hat{\mathcal A}}^{\dag}(t), \\ 
{\cal T}_s \protect{\hat{\mathcal A}}(t) \protect{\hat{\mathcal P}}^{(s)}_{>t} = 
\frac{1-s}{2}\protect{\hat{\mathcal A}}(t)\protect{\hat{\mathcal P}}^{(s)}_{>t} + 
\frac{1+s}{2}\protect{\hat{\mathcal P}}^{(s)}_{>t}\protect{\hat{\mathcal A}}(t)
, 
 \\ 
{\cal T}_s \protect{\hat{\mathcal A}}^{\dag}(t) \protect{\hat{\mathcal P}}^{(s)}_{>t} = 
\frac{1+s}{2}\protect{\hat{\mathcal A}}^{\dag}(t)\protect{\hat{\mathcal P}}^{(s)}_{>t} + 
\frac{1-s}{2}\protect{\hat{\mathcal P}}^{(s)}_{>t}\protect{\hat{\mathcal A}}^{\dag}(t)
, 
\end{gathered}}}
% \nonumber % \eqlabel{} 
\end{gather}}%
%+++++++++++++++++++++++++++++++++++++++++++++
where $\protect{\hat{\mathcal P}}^{(s)}_{>t}$ is a time-$s$-ordered product of operators with all time arguments larger than $t$. 

Properties: 
\begin{itemize}
\item
For $s=1$ and $s=0$: time-normal and time-symmetric products.
\item
For other $s$---new orderings; e.g., time-antinormal for $s=-1$.
\item
Hermitian-conjugate of a ${\cal T}_s$-product is a ${\cal T}_s$-product. 
\item
Continuity? (hopefully)
\end{itemize}
%*******************************************
%\begin{figure} 
%\begin{center} 
%\includegraphics[width=0.99\columnwidth]{X.eps}
%\end{center} 
%\caption{XXX} 
%\end{figure} 
%*******************************************
%\begin{acknowledgments}
%*******************************************
%This research was supported by the Marsden Fund of the Royal 
%Society of New Zealand, the New Zealand Foundation for Research, 
%Science and Technology 
%(UFRJ0001) and the Deutsche Forschungsgemeinschaft under 
%contract FL210/11.
%*******************************************
%\end{acknowledgments}
%******************************************* 
\appendix
%******************************************* 
\begin{widetext} 
%******************************************* 
\section{Infinitesimal transition amplitude in the $S$-representation} 
\label{app:InfUS}
%******************************************* 
Here we outline the reasoning leading from eq.\ (\protect\ref{eq:UInfS}) to (\ref{eq:UInfS1}). We assume that the differential operator in (\ref{eq:UInfS}) may be drawn under the integral and applied directly to the two factors comprising the integrand. Applying 
$\exp\protect\ensuremath{\big(
\frac{s}{2} \,{
{ 
{\frac{\partial^2 }
{\partial\alpha \partial\alpha^* }}
}} \big)}$ 
is straightforward and results in a factor $\exp\protect\ensuremath{\big(
-\frac{s|\sigma |^2}{2} \big)}$. Applying 
$\exp\protect\ensuremath{\big(
-\frac{s}{2} \,{
{ 
{\frac{\partial^2 }
{\partial\alpha_0 \partial\alpha_0^* }}
}} \big)}$ 
is also easy if using the formula (with $F_1$, $F_2$ and $F_3$ being arbitrary complex functions)
%=============================================
\protect{\begin{align}{{
 \begin{aligned} 
F_1\protect\ensuremath{\bigg(
\frac{\partial }{\partial \alpha_0},
\frac{\partial }{\partial \alpha_0^*}
 \bigg)} F_2(\alpha_0)F_3(\alpha_0) %\\ 
= 
F_1\protect\ensuremath{\bigg(
\frac{\partial }{\partial \alpha_0} + 
\frac{\partial }{\partial \alpha_0'},
\frac{\partial }{\partial \alpha_0^*} + 
\frac{\partial }{\partial \alpha_0^{\prime *}}
 \bigg)} F_2(\alpha_0)F_3(\alpha_0') \big|_{\alpha_0'=\alpha_0}, 
\end{aligned}}}
% \nonumber % \eqlabel{} 
\end{align}}%
%+++++++++++++++++++++++++++++++++++++++++++++
which expresses the rule of product differentiation. 
To start with, we write 
%=============================================
{\begin{multline} 
\exp\protect\protect\ensuremath{\bigg[
-\frac{s}{2} \,{
{ 
\protect\ensuremath{\Big(
{\frac{\partial } 
{\partial\alpha_0}} + 
{\frac{\partial } 
{\partial\alpha'_0}}
 \Big)} 
\protect\ensuremath{\Big(
{\frac{\partial } 
{\partial\alpha^*_0}} + 
{\frac{\partial } 
{\partial\alpha^{\prime *}_0}}
 \Big)} 
}}\bigg]} 
\\ = 
\exp\protect\protect\ensuremath{\bigg[
-\frac{s}{2} \,{
{ 
\protect\ensuremath{\Big(
{\frac{\partial^2 }{\partial\alpha_0 \partial\alpha_0^{\prime *} } + 
{\frac{\partial^2 }{\partial\alpha'_0 \partial\alpha_0^{*} } 
}} \Big)}}}\bigg]} 
\exp\protect\ensuremath{\bigg(
-\frac{s}{2} \,{
{ 
\frac{\partial^2 }{\partial\alpha_0 \partial\alpha_0^* }
}}
 \bigg)}
\exp\protect\ensuremath{\bigg(
-\frac{s}{2} \,{
{ 
{\frac{\partial^2 }{\partial\alpha'_0 \partial\alpha_0^{\prime *} }}
}} \bigg)} . 
% \nonumber % \eqlabel{} 
\end{multline}}%
%+++++++++++++++++++++++++++++++++++++++++++++
Then, firstly, 
%=============================================
\protect{\begin{align}{{
 \begin{aligned} 
\exp\protect\ensuremath{\bigg(
-\frac{s}{2} \,{
{ 
{\frac{\partial^2 }
{\partial\alpha_0 \partial\alpha_0^* }}
}} \bigg)}\exp\protect\protect\ensuremath{\big[
(\alpha -\alpha _0)\sigma ^* - (\alpha -\alpha _0)^*\sigma
\big]} 
= \exp\protect\protect\ensuremath{\big[
(\alpha -\alpha _0)\sigma ^* - (\alpha -\alpha _0)^*\sigma + s|\sigma |^2/2
\big]} . 
\end{aligned}}}
% \nonumber % \eqlabel{} 
\end{align}}%
%+++++++++++++++++++++++++++++++++++++++++++++
This cancels $\exp\protect\ensuremath{\big(
-\frac{s|\sigma |^2}{2} \big)}$ emerging from $\exp\protect\ensuremath{\big(
\frac{s}{2} \,{
{ 
{\frac{\partial^2 }
{\partial\alpha \partial\alpha^* }}
}} \big)}$. Secondly, from eq.\ (\protect\ref{eq:EqS})
%=============================================
{\begin{multline} 
\exp\protect\ensuremath{\bigg(
-\frac{s}{2} \,{
{ 
{\frac{\partial^2 }
{\partial\alpha'_0 \partial\alpha_0^{\prime *} }}
}} \bigg)}\protect\protect\ensuremath{\bigg\{
1 - \frac{i\Delta t}{\hbar } \protect\protect\ensuremath{\Big[
H(\alpha'_0 - \sigma /2,t) - H(\alpha'_0 + \sigma /2,t)
\Big]}
\bigg\}} \\ 
= 1 - \frac{i\Delta t}{\hbar } \protect\protect\ensuremath{\Big[
H_s(\alpha'_0 - \sigma /2,t) - H_s(\alpha'_0 + \sigma /2,t)
\Big]} .
% \nonumber % \eqlabel{} 
\end{multline}}%
%+++++++++++++++++++++++++++++++++++++++++++++
Thirdly, by using the commutational relations for derivatives, 
%=============================================
\protect{\begin{align}{{
 \begin{aligned} 
\frac{\partial }{\partial \alpha_0}
\exp\protect\protect\ensuremath{\big[
(\alpha -\alpha _0)\sigma ^* - (\alpha -\alpha _0)^*\sigma
\big]} & = \exp\protect\protect\ensuremath{\big[
(\alpha -\alpha _0)\sigma ^* - (\alpha -\alpha _0)^*\sigma
\big]}\protect\ensuremath{\bigg(
\frac{\partial }{\partial \alpha_0} - \sigma ^*
 \bigg)} , 
\\ 
\frac{\partial }{\partial \alpha^*_0}
\exp\protect\protect\ensuremath{\big[
(\alpha -\alpha _0)\sigma ^* - (\alpha -\alpha _0)^*\sigma
\big]} & = \exp\protect\protect\ensuremath{\big[
(\alpha -\alpha _0)\sigma ^* - (\alpha -\alpha _0)^*\sigma
\big]}\protect\ensuremath{\bigg(
\frac{\partial }{\partial \alpha^*_0} + \sigma 
 \bigg)} , 
\end{aligned}}}
% \nonumber % \eqlabel{} 
\end{align}}%
%+++++++++++++++++++++++++++++++++++++++++++++
we have 
 %\hfill
%=============================================
\protect{\begin{gather}{{
 \begin{gathered} 
\exp\protect\protect\ensuremath{\bigg[
-\frac{s}{2} \,{
{ 
\protect\ensuremath{\Big(
{\frac{\partial^2 }{\partial\alpha_0 \partial\alpha_0^{\prime *} } + 
{\frac{\partial^2 }{\partial\alpha'_0 \partial\alpha_0^{*} } 
}} \Big)}}}\bigg]} 
\exp\protect\protect\ensuremath{\big[
(\alpha -\alpha _0)\sigma ^* - (\alpha -\alpha _0)^*\sigma
\big]}
\hfill \\ \hfill\times
\protect\protect\ensuremath{\bigg\{
1 - \frac{i\Delta t}{\hbar } \protect\protect\ensuremath{\Big[
H_s(\alpha'_0 - \sigma /2,t) - H_s(\alpha'_0 + \sigma /2,t)
\Big]}
\bigg\}}\big|_{\alpha'_0 = \alpha_0} \\ 
\hspace{0.025\textwidth} = \exp\protect\protect\ensuremath{\big[
(\alpha -\alpha _0)\sigma ^* - (\alpha -\alpha _0)^*\sigma
\big]}
\exp\protect\ensuremath{\bigg(
\frac{s\sigma ^*}{2}{\frac{\partial }{\partial\alpha_0^{*} } - 
\frac{s\sigma }{2}{\frac{\partial }{\partial\alpha_0} 
}} \bigg)} 
\hfill \\ \hfill\times
\protect\protect\ensuremath{\bigg\{
1 - \frac{i\Delta t}{\hbar } \protect\protect\ensuremath{\Big[
H_s(\alpha_0 - \sigma /2,t) - H_s(\alpha_0 + \sigma /2,t)
\Big]}
\bigg\}}\\ 
\hspace{0.025\textwidth} = \exp\protect\protect\ensuremath{\big[
(\alpha -\alpha _0)\sigma ^* - (\alpha -\alpha _0)^*\sigma
\big]}
\protect\protect\ensuremath{\bigg\{
1 - \frac{i\Delta t}{\hbar } \protect\protect\ensuremath{\bigg[
H_s\protect\ensuremath{\Big(
\alpha_0 - \frac{(1+s)\sigma }{2},
\alpha^*_0 - \frac{(1-s)\sigma^*}{2} ,t
 \Big)} 
\hfill \\ \hfill -
H_s\protect\ensuremath{\Big(
\alpha_0 + \frac{(1-s)\sigma}{2},
\alpha^*_0 + \frac{(1+s)\sigma^*}{2},t
 \Big)} 
\bigg]}
\bigg\}} . 
\end{gathered}}}
% \nonumber % \eqlabel{} 
\end{gather}}%
%+++++++++++++++++++++++++++++++++++++++++++++
In this formula, we had to amend our notation, 
%=============================================
\protect{\begin{align}{{
 \begin{aligned} 
H_s(\alpha,t)\to H_s(\alpha,\alpha^*,t), 
\end{aligned}}}
% \nonumber % \eqlabel{} 
\end{align}}%
%+++++++++++++++++++++++++++++++++++++++++++++
to accommodate for asymmetry between $\alpha$ and $\alpha^*$ introduced by 
a non-symmetric ordering. We have thus indeed recovered the integrand of eq.\ (\protect\ref{eq:UInfS1}). 
\end{widetext}%
%******************************************* 
 
%*******************************************

\begin{thebibliography}{9}
\bibitem{But2}
A meaningful discussion of to what extent such deterministic trajectory in phase space is a physically classical object must include analyses of its relation to measurements, and would take us far outside the scope of this paper.
\bibitem{But}
Such clear separation of the deterministic motion from quantum fluctuations is only characteristic of the Wigner representation where the second derivative of $H(\alpha ,t)$ does not contribute. In principle, noises due to quadratic terms may also be classical (linear or nonlinear diffusion around the deterministic trajectory).
\bibitem{But3}
This inference is subject to continuity of the paths, to be proven elsewhere.
\end{thebibliography}
\end{document}